# Diverse dust populations in the near-Sun environment characterized by PSP/ISɸIS

M. M. Shen[1], J. R. Szalay[1], P. Pokorný[2,3], J. G. Mitchell[4], M. E. Hill[5], D. G. Mitchell[5], D. J. McComas[1], E. R. Christian[6], C. M. S. Cohen[7], N. A. Schwadron[8], S. D. Bale[9,10] and D. M. Malaspina[11,12]

1. *Department of Astrophysical Sciences, Princeton University, Princeton, NJ 08544*
2. *Astrophysics Science Division, NASA Goddard Spaceflight Center, Greenbelt, MD 20771*
3. *Department of Physics, The Catholic University of America, Washington, DC 20064*
4. *Heliospheric Science Division, NASA Goddard Spaceflight Center, Greenbelt, MD 20771*
5. *Johns Hopkins University Applied Physics Laboratory, Laurel, MD 20723*
6. *Goddard Space Flight Center, Greenbelt, MD 20771*
7. *California Institute of Technology, Pasadena, CA 91125*
8. *University of New Hampshire, Durham, NH 03824*
9. *Space Sciences Laboratory, University of California, Berkeley, CA 94720*
10. *Physics Department, University of California, Berkeley, CA 94720*
11. *Laboratory for Atmospheric and Space Physics, University of Colorado, Boulder, CO 80303*
12. *Department of Astrophysics and Planetary Sciences, University of Colorado, Boulder, CO 80309*

## Key Points (if applicable):

(1) ISɸIS/EPI-Lo has registered eight direct dust puncture events over PSP's first twenty orbits near the Sun, and several were the closest direct dust detections to the Sun.

(2) Investigating dust puncture events helps constrain the existing meteoroid model by providing location, flux, speed, and source category information.

(3) One of these impactors was likely from a retrograde dust grain, suggesting long-period cometary material may survive within 0.095 au (20.5 Rs).

Corresponding author: mitchellshen@princeton.edu

# Abstract

The Integrated Science Investigation of the Sun (IS☉IS) energetic particle instrument suite on Parker Solar Probe is dedicated to measuring energetic ions and electrons in the near-Sun environment. It includes a half-sky-viewing time-of-flight mass spectrometer (EPI-Lo) and five high-energy silicon solid-state detector-telescopes (EPI-Hi). To August 2024, eight of EPI-Lo's eighty separate telescope foils have experienced direct dust puncture events, most of which occurred inside 40 solar radii (0.19 au). These impacts represent the closest ever direct dust detections to the Sun. While there is limited information about the size/mass of each impact due to the lack of a dedicated dust instrument, we can determine the impact direction for six punctures, allowing us to partially constrain the inner zodiacal abundance. Remarkably, one of six unambiguous dust impacters was likely on a retrograde orbit, suggesting long-period cometary material may survive within 20 solar radii (0.09 au). We discuss observations in the context of improving our understanding of the inner zodiacal dust environment, highlighting multiple dust populations responsible for these events, and refining hazard assessment for near-Sun spacecraft.

# 1. Introduction

Parker Solar Probe (PSP) was launched in August 2018 with the primary objective of exploring the very inner heliospheric environment for the first time. With Venus gravity assists, PSP has gradually decreased its perihelion distances, eventually reaching altitudes less than 10 solar radii ($R_S$) at the closest approach. The goals of the Parker Solar Probe mission are (i) to investigate the coronal heating mechanism and the release of the solar wind, (ii) to investigate the solar wind plasma structures and how they relate to the features in the interplanetary magnetic field, and (iii) to study the acceleration and transport of energetic particles (*Fox et al.*, 2016).

PSP carries four scientific instrument suites: Integrated Science Investigation of the Sun (ISⵙIS – *McComas et al.*, 2016), Electromagnetic Fields Investigation (FIELDS – *Bale et al.*, 2016), Solar Wind Electrons Alphas and Protons (SWEAP – *Kasper et al.*, 2016) investigation, and Wide-field Imager for Solar Probe (WISPR – *Vourlidas et al.*, 2016). WISPR is a white light imager that investigates coronal structures and interplanetary transients, SWEAP measures the properties of solar wind plasma, and FIELDS characterizes the electromagnetic environment of the inner heliosphere. ISⵙIS, the primary topic of this work, is a suite of energetic particle instruments designed to provide measurements of solar energetic particles (SEPs) closer to the Sun than ever before (*McComas et al.*, 2019). ISⵙIS offers a novel opportunity to study particle acceleration at the Sun with minimal contribution from particle transport effects. In addition to SEP investigations, ISⵙIS data has been used to study cosmic rays (*Rankin et al.*, 2021; 2022), particles accelerated by stream interaction regions (e.g., *Cohen et al.*, 2020; *Allen et al.*, 2020; *Desai et al.*, 2020), solar $\gamma$-ray (*Mitchell et al.*, 2024), Jovian electrons (*Mitchell et al.*, 2022), etc. This study extends the scientific range of PSP/ISⵙIS data to examine the abundant interplanetary dust environment of the very inner heliosphere.

Interplanetary dust populations carry signatures of their parent bodies and the complex dynamical interactions that shape their orbital trajectories over time. The primary sources of these dust grains are comets, asteroids, and Edgeworth-Kuiper objects (e.g., *Nesvorný et al.*, 2010; *Poppe*, 2016; *Pokorný et al.*, 2024). These dust grains, gravitationally bound to the Sun, undergo an orbital transformation from elliptical to more circular trajectories as they spiral towards the Sun. A gradual loss of angular momentum is exerted by Poynting–Robertson (P-R) drag related to solar radiation pressure (*Burns et al.*, 1979). As these dust grains approach the Sun, their spatial density increases substantially, leading to collision fragmentation (*Pokorný et al.*, 2024) and sublimation into smaller grains (*Mann et al.*, 2004). The most susceptible grains to the outward force of solar radiation pressure are those with sub-micron sizes (radii on the order of 100s nm). Their orbital characteristics are determined by a parameter known as $\beta$, representing the ratio of solar radiation pressure to gravitational force, i.e., $\beta = F_R/F_G$ (*Burns et al.*, 1979). This value depends on both the size and composition of the grains. Grains with $\beta$ values exceeding a critical threshold will have positive orbital energy, namely $\beta$–meteoroids, following hyperbolic trajectories to escape from the heliosphere. In contrast, before such escape, all gravitationally bounded grains are categorized as $\alpha$–meteoroids (*Sommer*, 2023).

Multiple spacecraft have observed the presence of escaping $\beta$–meteoroids in our solar system, for instance, Pioneers 8 and 9 (*Berg & Grun*, 1973; *Grun et al.*, 1985), Helios (*Grun et al.*, 1980), Ulysses (*Wehry & Mann*, 1999), STEREO (*Zaslavsky et al.*, 2012), PSP (*Szalay et al.*, 2020; 2021; *Page* et al., 2020; *Malaspina* et al., 2020), and Solar Orbiter (*Zaslavsky et al.*, 2021). Electromagnetic forces play a significant role in guiding the paths of nanograins (typically ≲ 50 nm); however, they have minimal impact on the dynamics of larger grains in the inner solar system (*Morfill et al.*, 1986), and modern instruments are less sensitive to this size range (*Meyer-Vernet et al.*, 2014). Overall, dust grains are continually generated and possess finite lifetimes shaped by gravitational, solar radiation pressure, and electromagnetic forces.

Collisions play a significant role in eroding the zodiacal cloud, followed by either expelling large amounts of materials from the heliosphere, primarily through the escape of $\beta$–meteoroids (*Grun et al.*, 1985) or producing pick-up ions as grains P-R drag into the very inner heliosphere and sublimate (*Wimmer-Schweingruber and Bochsler*, 2003; *Schwadron and Gloeckler*, 2007). Sublimation and sputtering rates are estimated to be orders of magnitude lower than collisions (*Mann & Kimura*, 2000). Since $\beta$–meteoroid escape is the most efficient mechanism for the zodiacal cloud's material loss, it directly reflects the intensity of collisional processing within the inner heliosphere (*Grun et al.*, 1985; *Pokorný et al.*, 2024).

The zodiacal cloud comprises two primary populations of gravitationally bound $\alpha$–meteoroid grains: (a) azimuthally symmetric background and (b) discrete meteoroid streams. Our understanding of meteoroid streams relies on meteor shower observations occurring when Earth intersects them (e.g., *Jenniskens*, 1994); however, most streams do not intersect Earth (*Soja et al.*, 2019). Additionally, not all streams are consistent with generation via cometary activity. For instance, the Geminids meteoroid stream, one of the Earth's most intense, links to the active asteroid (3200) Phaethon but lacks the activity needed to maintain its stream (*Jewitt et al.*, 2013). As PSP sweeps near Phaethon orbits and/or other asteroids and comets, it may be detecting dust impacts from their streams (*Szalay et al.*, 2021; *Pusack et al.*, 2021; *Cukier & Szalay*, 2023). Hence, PSP has become a valuable tool for exploring previously uncharted regions of the zodiacal cloud, particularly in the near Sun dust environment.

Prior to this study of PSP/ISℨIS dust detections through puncture events, most studies used the PSP/FIELDS antenna instrument (*Bale et al.*, 2016) which detects dust impact on the spacecraft surface through impact ionization (e.g., *Malaspina et al.*, 2020, 2023; *Mozer et al.*, 2020; *Page et al.*, 2020; *Pusack et al.*, 2021; *Szalay et al.*, 2020, 2021). In impact ionization, the initial kinetic energy of the particle happens to heat, evaporate, and ionize the particle itself with the target material when impact speeds are $\gtrsim 1$ km/s (*Auer*, 2001). A plasma cloud will be generated upon impact and interact with the spacecraft and antennae, generating measurable voltage signals (see details in *Shen et al.*, 2021a; 2021b). Impact ionization makes antenna-based instruments sensitive to interplanetary and interstellar dust impacts with submicron to micron sizes (*Mann et al.*, 2004), making it commonly applied in missions. However, it has limitations compared to dedicated dust detection as it is difficult to (a) adequately discriminate dust impact signals from plasma wave measurements, (b) convert from measured voltage to impact charge to mass/size, and (c) account for the ambient plasma environment effect (*Shen et al.*, 2021b; 2023).

Once high-speed grains impact the foils of an energetic particle instrument, the collimator/detection foils can be punctured, leading to irreversible damage and enhanced photon background contamination to the measurements. PSP/ISℨIS has encountered eight dust punctures, most occurring within 40 R$_S$ — the closest direct dust detections to the Sun. Although size/mass details remain limited, we have determined six impact source directions, enabling us to constrain the abundance of material in the inner zodiacal cloud. Early predictions estimated a cumulative impact count of 0.4 – 3.0 over the first two orbits on an example EPI-Lo's collimator foil (single aperture L31 analysis in *Szalay et al.*, 2020), and ISℨIS/EPI-Lo has registered eight punctures over the first twenty orbits from August 2018 to August 2024. Given the limitations of the antenna-based dust dataset and lack of direct dust impact measurements in the near Sun environment, we apply a three-dimensional dynamical zodiacal dust model to six unambiguous dust puncture measurements collected by ISℨIS for data-model comparison.

# 2. ISƟIS Energetic Particle Instrument

The Integrated Science Investigation of the Sun (ISƟIS) suite comprises the low- and high-energy Energetic Particle sensors, EPI-Lo and EPI-Hi, respectively (*McComas et al.,* 2016), measuring energetic ions in the range of 20 keV/nucleon – 200 MeV/nucleon and electrons from ~ 25 keV – 6 MeV. EPI-Lo is a time-of-flight (TOF) mass spectrometer that utilizes TOF within the instrument in concert with energy deposit in a solid-state detector (SSD) to determine the mass and energy of incident ions (*Hill et al.*, 2017). EPI-Lo is divided into eight instrumental "wedges," each of which has ten individual look apertures, as shown in Fig.1 (a). EPI-Lo's eighty apertures across eight wedges sample approximately hemispherical sky coverage, enabling the ability to measure detailed anisotropy of energetic particles.

Each aperture of EPI-Lo includes a collimator with an outer baffle to define the field of view (FOV) and to mitigate the risk of dust impact and light contamination, as Fig. 1 (b) depicts. To prevent spurious signals generated by ambient solar wind electrons or photons into the instrument, a combination of two foils significantly reduces the effect of pinholes produced by dust punctures. The first thin polyimide foil (collimator foil) is placed midway down each collimator to protect the instrument further while only minimally affecting the measured energy of incident ions. A second foil (start foil) lies at the junction between each aperture and the wedge body, primarily for releasing secondary electrons in response to the passage of incident ions for TOF measurements. On EPI-Lo, the six most Sun-looking apertures require additional light protection and have a combination of 0.11 μm thicker layers on start foils to mitigate scattered photon contamination (*Hill et al.*, 2017).

The method of ion detection of ISƟIS/EPI-Lo is described in detail by *McComas et al.* (2016) and *Hill et al.* (2017). An incident ion enters the instrument through one of the apertures and knocks an electron free from the start foil. This electron is then steered toward a microchannel plate (MCP) detector at the bottom of each wedge, producing a start signal. A mask covering the MCP has openings corresponding to each aperture such that the location of start signal on MCP allows the identification of which aperture the ion enters. The ion then travels to the rear of the wedge, interacts with a stop foil, and releases another secondary electron. This electron is steered with an electrostatic mirror to another opening in the MCP mask, producing a stop signal. Combining the start and stop signals produces the ion TOF measurements. The ion then deposits its remaining energy on an SSD at the rear of the wedge. Between the TOF measurement and the energy-deposit measurement, ISⓄIS/EPI-Lo provides high-resolution energy, mass, and spatial measurements.

Throughout the first twenty orbits of PSP mission, ISⓄIS/EPI-Lo experienced eight dust hits. These dust holes have resulted in spurious start signals or potential stop signals depending on the trajectory of the ambient contaminating photons, thus triggering accidental coincidences in the TOF-only data (i.e., events that require a TOF measurement only with no required energy deposited on the SSD). On the other hand, the "triple-coincidence" data products, requiring a start and a stop on MCP along with a signal on SSD, are not significantly affected by these accidental coincidences.

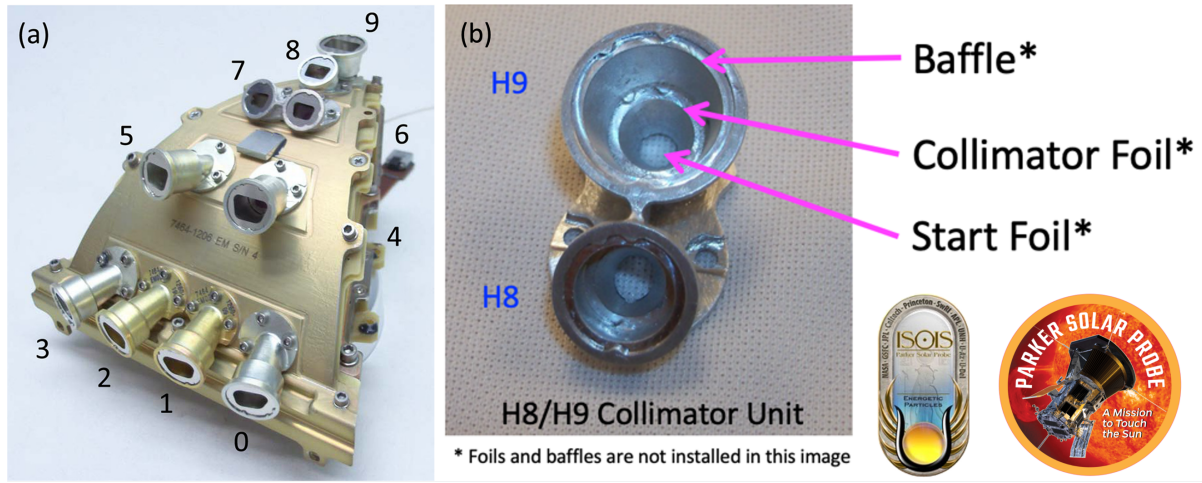

**Figure 1:** Instrumental design of ISⓄIS/EPI-Lo suite. Panel (a) depicts ten individual apertures of each wedge. Panel (b) shows the baffle and collimator design before the start foil.

A dust strike in an EPI-Lo aperture is identified when evidence of an instantaneously appearing hole in the EPI-Lo foil (or foils) suddenly increases the transmitted light and, thus, detector counts. Locally, upon a dust hit, the signal is typified by an instantaneous jump when the instrument is in a bright environment (i.e., close to the Sun), where the term "instantaneous" is limited by the measurement cadence only (per second on EPI-Lo). Depending on local conditions, it is not always possible to identify the time of impact as the elevated noise background from the puncture may not be sufficiently high until near the Sun in orbit. In most cases, the signal jumps from a background level of counts to a distinctly higher rate during a single measurement interval. Moreover, the scattered photon contamination has an orbital and attitude dependence consistent with the known environment, e.g., brighter when closer to or oriented towards the Sun. Once an instantaneously appearing signal in a single aperture is identified with long-term behavior (orbital/attitude), a dust puncture event is confirmed. As a crosscheck, the energy response spectrum is used to identify scattered photon contamination to protect against being misled by some atypical particle measurements.

When a dust puncture event is confirmed, we compare the orientation of the foil surface upon impact with predictions from our understanding of dust populations in the inner heliosphere; this is especially challenging when the instrument is in a dark environment or orientation. So far, the dust punctures have generally appeared at times and locations with locally elevated probability for dust hits. If dust hits appear at unexpected times, the spacecraft's attitude is likely altered for specific purposes, typically for communications. For example, PSP often performs rotational maneuvers outside encounter periods (when PSP is within 0.25 au of the Sun). Depending on the exact orientation of the spacecraft relative to the propagating dust in that region, this may put EPI-Lo at an increased risk of dust impacts (e.g., the L35 event described later).

The geometric factor of the hole(s) made by the putative dust particle of a given size can be compared with the magnitude of the signal jump. In this way, a rough measure of the size or velocity of the dust grain may be estimated by modeling the amount of light allowed through the new hole. Further, such size or velocity information can be compared with an inner heliospheric dust model to seek consistency. Most of the time, only PSP/FIELDS antennas are sensitive to dust impacts because their effective collection area includes the entire spacecraft body and antennae surfaces (*Szalay et al.*, 2020; *Malaspina et al.*, 2020; *Szalay et al.*, 2021; *Pusack et al.*, 2021). For five of the eight definitive impacts, we can link the impact waveforms registered by the FIELDS instrument to the actual dust hits collected by the ISOIS energetic particle suite as "correlated" dust identification among spacecraft subsystems (see Appendix).

There are multiple ways that dust impacts can damage the ISOIS foils and produce a detectable increase in backgrounds. A particularly large dust puncture (L28 event in orbit 9) prompted us to examine mechanisms for producing holes on foils beyond direct hits. Figure 2 sketches the possible impact scenarios. Scenario 2 depicts a simple puncture produced by incident dust penetrating both the collimator foil (blue) and the start foil (green). Scenario 3 illustrates a dust particle narrowly missing the collimator foil, impacting the holder for the collimator foil and producing further debris upon impact. Hence, the debris neither punctured the foil nor made it all the way through. Scenario 1, instead, presents a situation that can pose a risk of significant background enhancement. Here, the incident dust grain impacts the inner surface of the collimator turret outside the collimator foil with sufficient momentum, knocking aluminum debris from the collimator body (cratered) that may punch larger holes in both collimator and start foils than that just a direct hit. This large aluminum debris may also become lodged in the instrument body, causing further issues than simply scattered photon contamination. Accordingly, the geometric factor that can produce such damage to the collimator and start foils is not merely the geometric factor of the collimator and outer baffle but also the geometric factor of all possible lines of sights from the inner face of the collimator body through both the collimator and start foils.

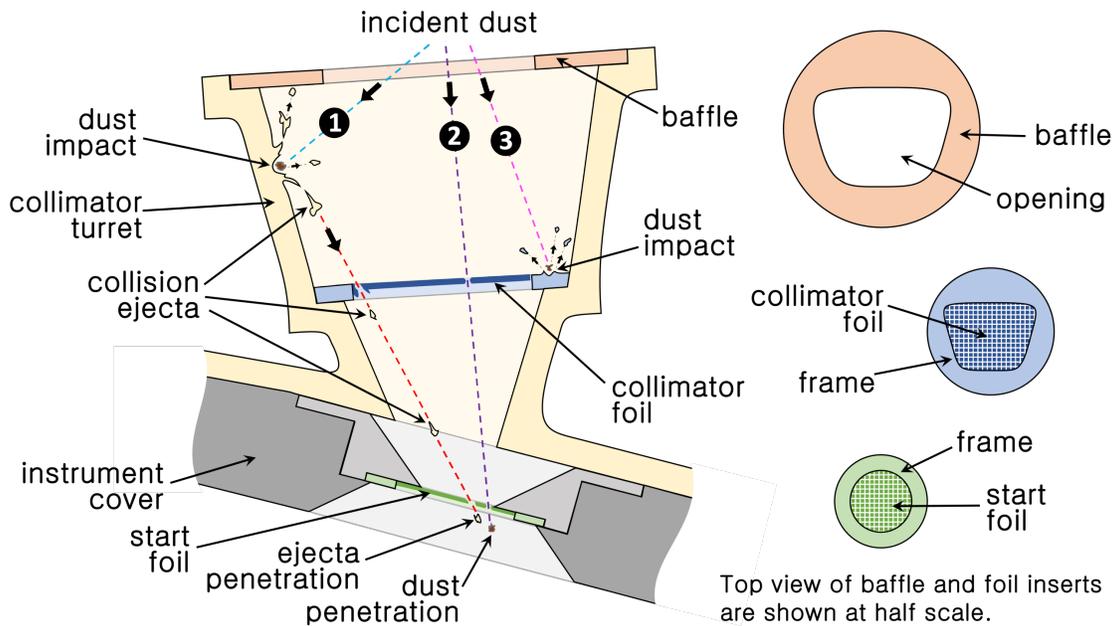

**Figure 2:** Schematic of three possible mechanisms of dust puncturing an ISʘIS/EPI-Lo channel aperture. The components in sketch are up to scale.

We implemented a Geant4 model (*Agostinelli et al.*, 2003) of EPI-Lo to quantify the surface area within each collimator that could produce a secondary projectile capable of piercing the foil and entering the instrument along a straight path. The six unique collimator designs on each EPI-Lo wedge necessarily have different probabilities of producing this type of event, as Table 1 listed, referring to aperture 0, 1, 4, 6, 8, 9 in Fig. 1, where four subsets of (0 & 3), (1 & 2), (4 & 5), and (6 & 7) are geometric symmetricity.

To characterize these regions, we initially traced 200 keV protons from a $2\pi$ hemispherical "cap" in the instrument's interior underneath the start foil directed toward the outside of the instrumental aperture (i.e., inner hits, 3rd column in Table 1). The locations of dust particle interactions between the outer baffle and the collimator foil from these simulations were compiled (4th column in Table 1), and the regions defined by these points were mathematically bounded using a convex hull algorithm. Next, we produced a hemispherical isotropic cap outside the instrument and directed it toward the interior. The geometric factor for particles that interact in the regions of interest defined by the convex hull was calculated for each aperture. A combined geometric factor (GF) of inner hits, baffle, and collimator is provided in the 5th column in Table 1. The resulting geometric scaling factors of individual apertures are then calculated by dividing these combined GFs by the 0.226 value initially presented in *Szalay et al.* (2020), which considered the effective opening area of collimator foil only.

**Table 1:** Geometric factor (GF) table of EPI-Lo apertures in unit of $cm^2$ sr with corresponding geometric scaling factors.

| EPI-Lo aperture | dust hits | GF of inner hits ($cm^2$ sr) | GF of baffle & collimator ($cm^2$ sr) | Combined GF ($cm^2$ sr) | Geometric scaling factor per aperture (unitless) |
|---|---|---|---|---|---|
| 0 & 3 | 0 | 0.380 | 0.249 | 0.629 | 2.783 |
| 1 & 2 | 2 | 0.460 | 0.275 | 0.735 | 3.252 |
| 4 & 5 | 3 | 0.840 | 0.102 | 0.942 | 4.168 |
| 6 & 7 | 1 | 0.018 | 0.307 | 0.325 | 1.438 |
| 8 | 1 | 0.120 | 0.211 | 0.331 | 1.465 |
| 9 | 1 | 0.550 | 0.181 | 0.731 | 3.235 |

The updated simulation results increased the overall geometric factor of potential dust hits in EPI-Lo by an averaged scaling factor of ~2.8x compared to the geometric factor calculated by *Szalay et al.* (2020). Such an increase in geometric factor brings the modeled dust hits provided by the dynamic impact flux model (Sec. 3) closer to what EPI-Lo experienced in the mission (Fig. 7) and explains enhanced dust hits on aperture (4 & 5) as well as (1 & 2) in Table 1.

# 3. Dynamic Impact Flux Model

To infer impactor categories from bound $\alpha$–meteoroids and unbound $\beta$–meteoroids that punctured ISƟIS/EPI-Lo look apertures, we perform a data-model comparison using a self-consistent collisional model that treats $\alpha$–meteoroids and their collisionally produced $\beta$–meteoroids as all part of the same process. It is built on the theoretical framework for understanding collisional evolution (grinding) and $\beta$–meteoroid production inside 1 AU (mass loss by expelling) (e.g., *Ishimoto and Mann*, 1998; *Mann et al.*, 2004; *Pokorný et al.*, 2024; *Szalay et al.*, 2021; *Zook and Berg*, 1975).

This model follows individual trajectories of many particles shed from four separate dust sources: Jupiter Family Comets (JFCs), Halley Type Comets (HTCs), Oort Cloud Comets (OCCs), and Main Belt Asteroids (MBAs) for grains with radii of 0.3 μm to 1 mm (*Pokorný et al.*, 2018; 2019; 2021; 2024). It is constrained by many observations, including the (a) total mass flux normalized consistent with the terrestrial mass flux, (b) orbital element distribution consistent with meteors observed at Earth, and (c) spatial number density consistent with Helios and STEREO observations. This two-component dynamical zodiacal dust model has also been employed successfully by data from the LADEE (Moon) and MESSENGER (Mercury) missions, making it robust in examining dust phenomena throughout the near Sun dust environment at least within 0.3 to 1.0 au.

The model inputs incorporate orbital (velocity) distribution, mass distribution, source populations, collision rates, and $\beta$–meteoroid production rates. The output parameters provide density, velocity, flux, and mass distributions. By integrating a substantial number of trajectories, we generate flux maps for any given position and velocity vector within the inner solar system. A Mollweide projection for the impact predictions has been used to visualize flux density distributions of bound $\alpha$–meteoroid categories in both prograde/retrograde orbits and spacecraft attitude along with the orbital ram direction of PSP (*Szalay et al.,* 2020). In model calculations, we provide the arrival directionality of unbound $\beta$–meteoroids with $\beta = 0.5$–1.2, following calculations described in *Szalay et al.* (2021). The vital benefit of this data-model comparison is we can constrain impactor categories by implementing this data-driven model. Results from the unambiguous ISƟIS/EPI-Lo direct events, including both inferred $\alpha$– and $\beta$–meteoroids puncture events, can be further referenced as the innermost dust impact sampling constraints.

# 4. Collected Puncture Events

## 4.1 Event Map of Dust Punctures

Figure 3 and Table 2 summarize the locations and look directions of eight dust impacts registered by ISOIS/EPI-Lo. For each puncture event, we label the orbit number followed by the impacted channel number. The first digit represents the wedge number of EPI-Lo (W0 – W7, Sec. 2), corresponding to the longitudinal information along the instrument's apex axis (e.g., W3 looks generally towards the Sun, while W7 sees generally anti-sunward). The second digit indicates the detector number (0 – 9), representing the look aperture (see Fig. 1 or *Hill et al.*, 2017). For example, L31 is detector 1 on wedge 3. Each impacted channel's field-of-view (FOV) cones are illustrated to help characterize the impacter categories and are highly determined by the spacecraft attitude/pointing. Notably, this FOV cone considers the instrument collimator foil only, which can be further enlarged to a broader open area by implementing the possibility of a dust grain impacting the inner baffles, as discussed in Sec. 2 and instrument paper (*Hill et al.*, 2017).

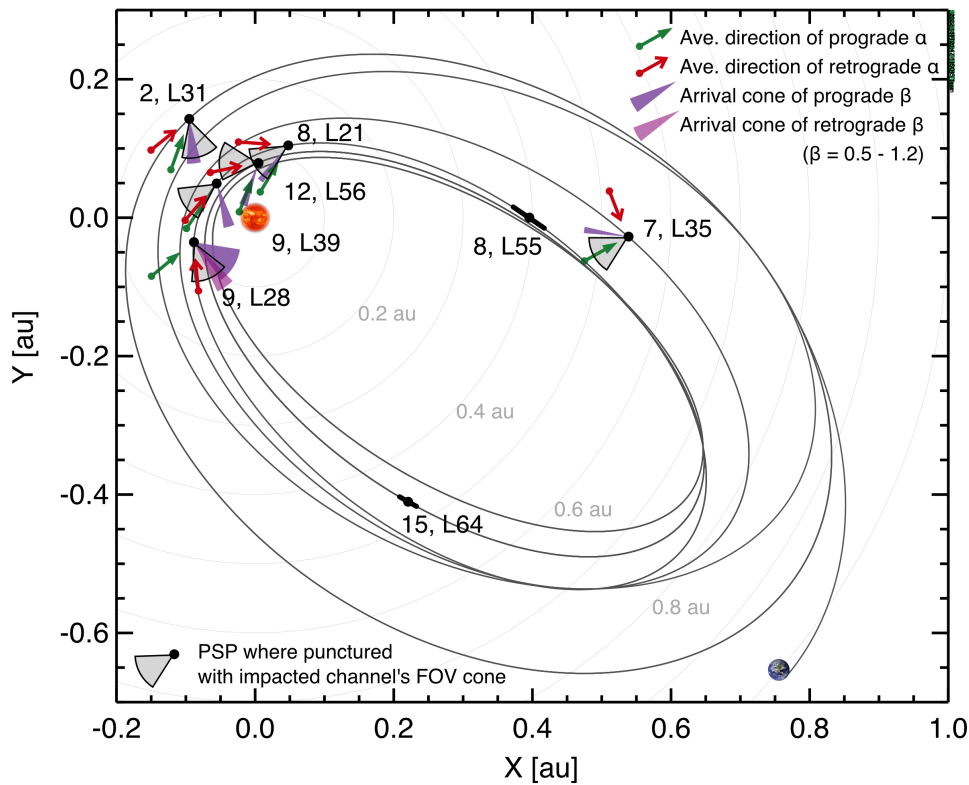

**Fig. 3:** Event map of dust punctures with corresponding FOV cone (grey sector) of the impacted detector channel (black dots) projected onto the J2000 ecliptic plane. The orbit numbers and affected channels are labeled nearby. The modeled source directions and categories of impacters

are indicated: average directions of bound $\alpha$–meteoroids are marked in arrows, while the shaded cones express $\beta$ = 0.5–1.2 of unbound $\beta$–meteoroids.

Figure 3 shows PSP's first twenty orbits. Employing the dust dynamical model (Sec. 3), both prograde and retrograde directions of bound $\alpha$–meteoroids on elliptic orbits and unbound $\beta$–meteoroids on hyperbolic orbits are projected to this two-dimensional impact map on the heliocentric J2000 ecliptic plane. To estimate the arrival cone of $\beta$–meteoroids (purple cones), the $\beta$ value is set to be 0.5 − 1.2 (*Szalay et al.*, 2021). This range of $\beta$ corresponds to a radius of ~0.2 − 0.6 μm with a bulk density of 2 g/cc (*Zook & Berg*, 1975).

To August 2024, ISⴲIS/EPI-Lo collected eight dust puncture events; five of eight were inside 0.2 au and on the pre-perihelion portion of the orbit. Inside 0.25 au, PSP maintains a fixed attitude and apex pointing. In orbits 8 and 15, the corresponding onset time of L55 and L64 events are marked as a dot with an uncertain time range since we cannot constrain it precisely while the instrument was powered-off. In addition, the channel L35 event in orbit 7 was an atypical case in which PSP spacecraft rotated 178 degrees along the sunward axis to point its high gain antenna towards Earth for communication, thus exposing ISⴲIS/EPI-Lo to a higher impact risk regime caused by typical, bound $\alpha$–meteoroids in prograde orbits.

## 4.2 Modeled Impact Flux and Impact Speed

Figure 4 shows the modeled impact flux of six unambiguous $\alpha$–meteoroid punctures on the exact impact time we determined. This flux map employs Mollweide projections for dust impact predictions on the entire PSP sky, where the center of the frame always points towards the Sun (heat shield sunward alignment), and the equatorial plane aligns with PSP's orbital plane with the ram direction is at ~90° in the right hemisphere. Each line represents 30° latitude or longitude. The arrival longitudes of simulated $\beta$–meteoroids with the assumptions described in *Szalay et al.* (2020) are labeled as a horizontal purple bar. The modeled impact speeds span from 0 to 400 km/s, where the average impact speed is shown in Fig. 5. The impacted channels are overlapped on the map with the detector center encircled by 25%, 50%, 75%, and 100% of FOV cones, respectively. These circles only represent the FOV cone of the instrument collimator foil. Through data-model comparison between depicted FOV cones and the featuring categories of impacters, we identify the most probable impactor source for each.

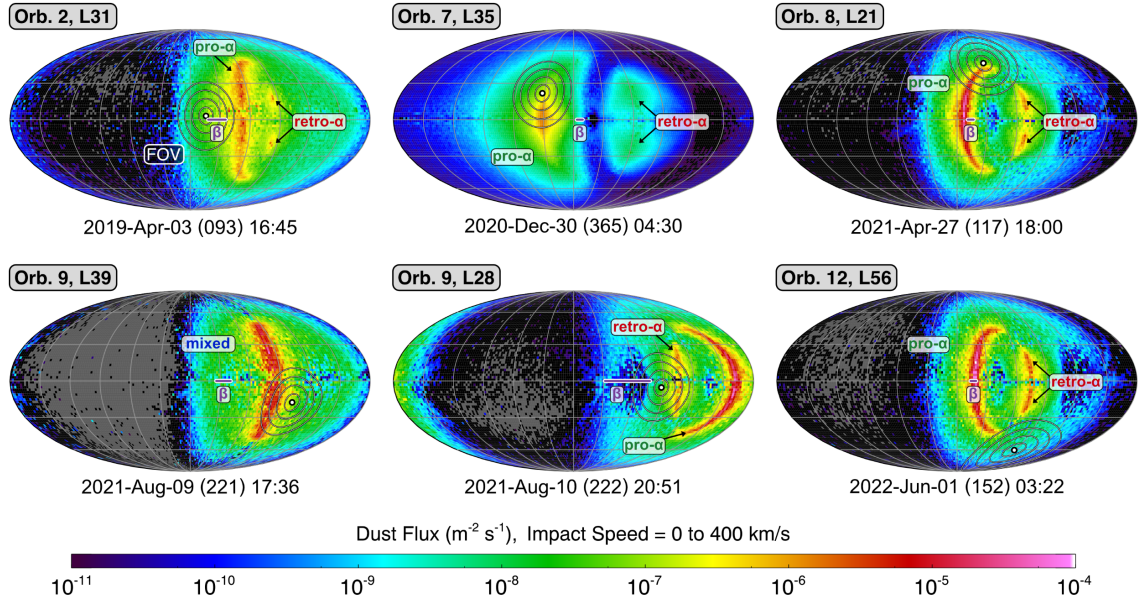

**Fig. 4:** Sky-map of modeled impact flux for six puncture events with impact speed spanning from 0 to 400 km/s.

Inside 1 au, interplanetary dust grains from Jupiter Family Comets (JFCs) dominate the number flux (*Nesvorný et al.,* 2010; *Poppe* 2016; *Pokorný & Kuchner* 2019; *Pokorný et al., 2024*). Most of the fluxes in these skymaps are from JFCs (which are all $\alpha$–meteoroids). Figure 4 shows two featured groups of $\alpha$–meteoroids in prograde and retrograde orbits. With the center of a skymap always sunward, the local dust flux enhancements are time-dependent as PSP's local position and velocity vector encounter different dust environments.

Impact speed serves as another key parameter to discriminate the potential impact categories (*Szalay et al.,* 2020; 2021). With foils as essential components of energetic particle detectors in space, the larger or faster the grain impacted, the more hazardous the consequence would be. In Fig. 5, the modeled average impact speed shown accounts for the spacecraft's orbital velocity vector. Similarly, the color scale only represents those of $\alpha$–meteoroids, while the $\beta$–meteoroids are labeled as horizontal purple bars. With most impacts occurring inside 0.2 au, the spacecraft has a larger specific angular momentum than the dust grains on circular orbits. Therefore, PSP "catches up" to grains on circular orbits in these regions; as such, the spacecraft frequently experienced impacts from the apex hemisphere, i.e., the right hemisphere of the skymap (*Szalay et al.,* 2020).

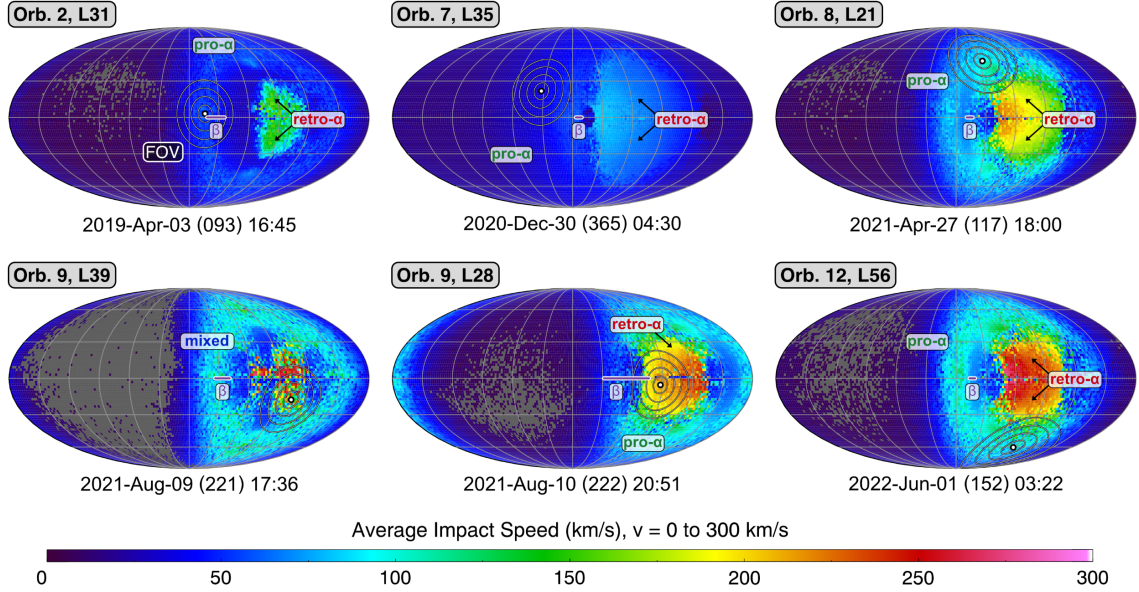

**Fig. 5:** Modeled impact speed across the entire sky in the same configuration as Figure 4.

In the first L31 event, both prograde–$\alpha$ and prograde–$\beta$ are possible. Considering the probability of averaging impact directions and up to 75% FOV doesn't overlap with the hot spots of retrograde–$\alpha$ (butterfly-like pattern), we suggest it was very likely to be punctured by a prograde $\beta$–meteoroid. Both L35 and L21 are clearly caused by prograde–$\alpha$ grains. Though $\alpha$–meteoroids in prograde or retrograde orbits can be distinguished in the map most of the time, event L39 in orbit 9 happened when two populations overlapped within the FOV, thus challenging the categorization. Considering prograde–$\alpha$ covers more of the FOV than retrograde–$\alpha$, we suggest it was likely caused by a prograde–$\alpha$ grain. Among these six events, five were likely to have relatively lower impact speeds of $\lesssim 100$ km/s. In contrast, event L28 in Orbit 9 significantly affected the instrument detector. Grains in "retrograde orbit" would have extremely high kinetic energy (as they would have impact speeds $\gtrsim 200$ km/s. As the L28 FOV overlaps with the retrograde-$\alpha$ peak flux locations, and given the severe damage from this impact, we infer this impactor was on a retrograde orbit. The FOV of L56 doesn't overlap with retrograde–$\alpha$ peak flux, suggesting the likelihood of another prograde–$\alpha$ impact event.

Table 2 summarizes all eight dust punctures with the corresponding orbit, impacted aperture, occurrence time, and the inferred impacter category via data-model comparison. Among these six events, L35, L21, L39, and L56 are recognized as prograde-$\alpha$ impacts. Event L31 was presumably punctured by a $\beta$–meteoroid, while that of L28 is most likely an $\alpha$ grain but in "retrograde orbit."

**Table 2:** Eight dust punctures with corresponding information and inferred population.

| Orbit | Aperture | Occurrence Time | Inferred Population |
|-------|----------|-----------------|---------------------|
| 2 | L31 | 2019–04–03 (093) 16:45 | prograde $\beta$ |
| 7 | L35 | 2020–12–30 (365) 04:30 | prograde $\alpha$ |
| 8 | L55 | 2021–04–17 (107) 16:45 to 2021–04–19 (109) 15:00 | N/A |
| 8 | L21 | 2021–04–27 (117) 18:00 | prograde $\alpha$ |
| 9 | L39 | 2021–08–09 (221) 17:36 | prograde $\alpha$ |
| 9 | L28 | 2021–08–10 (222) 20:51 | retrograde $\alpha$ |
| 12 | L56 | 2022–06–01 (152) 03:22 | prograde $\alpha$ |
| 15 | L64 | 2023–03–30 (089) 21:14 to 2023–04–01 (091) 02:08 | N/A |

## 4.3   Probability of Dust Impacts

Our dynamical model allows us to predict the instantaneous and cumulative impact rates across all eighty look apertures as a function of time, especially the directionality is well-determined. However, the flux and density of micron/sub-micron sized grains are not well-calibrated in our dynamical model, particularly near the Sun. Here, we use the ISΘIS/EPI-Lo detections to constrain the absolute abundances of small grains in the inner solar system. To do so, we calculate the total expected number of impacts to ISΘIS/EPI-Lo and compare it with the occurrence of observed impacts. We can then scale the model results to match the actual dust observations to approximate the abundances of inner solar system small dust grains. In Table 2 above, five $\alpha$–meteoroid impacts are identified, while the other two "N/A" ones are not possible to constrain the impactor origin with the current analyses given we do not know exactly when the punctures happened. Therefore, we determine the absolute scaling for the model to predict 5-7 $\alpha$–meteoroid impacts would have occurred to ISΘIS.

The probability of a dust impact with an average impact rate $\mu$ in time $\Delta t$ is given by $P(\mu, \Delta t) = 1 - e^{-\mu \Delta t}$. In a Monte-Carlo model, we generate 5000 instances of PSP flying through our simulated dust environment to estimate the total number of impacts received across all look directions. Figure 6 shows two resulting predictions for the total impacts to ISΘIS, with the model upscaled by factors of 1.42 and 1.85, respectively, to match the 5-7 impacts observed. We note that given the challenge of modeling such dust grains without previous measurements inside 0.3 au, it is remarkable the abundances that explain the ISƟIS impacts within less than a factor of 2.

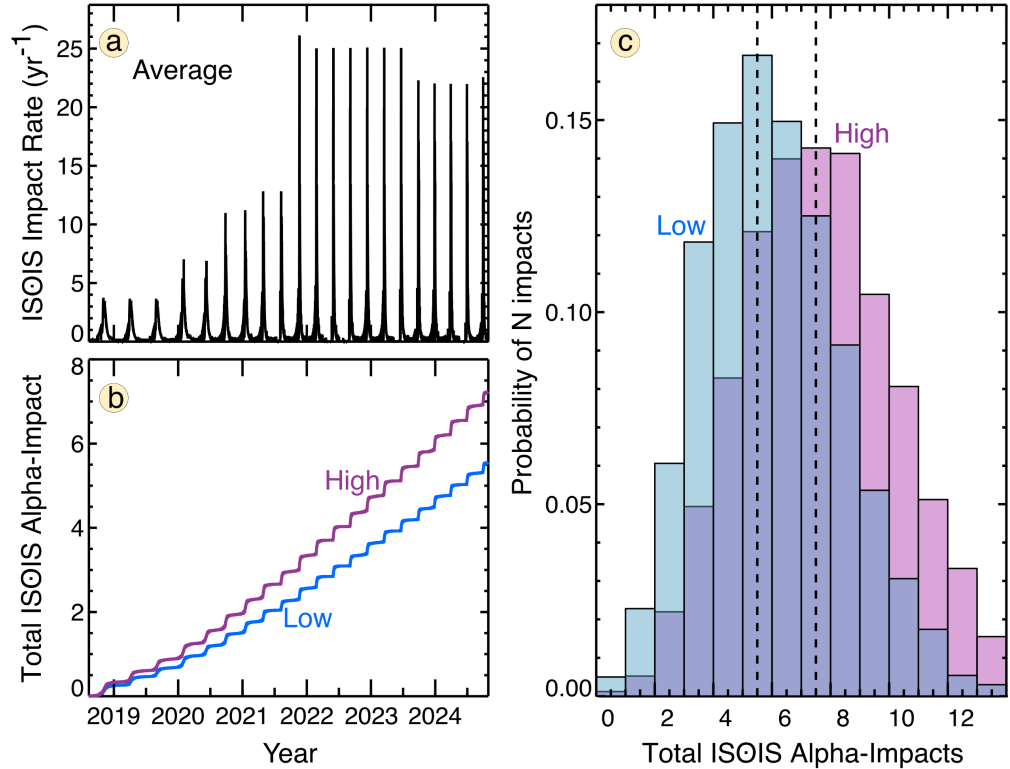

**Fig. 6:** Panel (a) provides modeled discrete impact rates over time. Panel (b) depicts the cumulative impacts over time with the probability of 5 (*Low*) and 7 (*High*) $\alpha$–meteoroid impacts, followed by the statistical scaling factors of 1.42 (*Low*) and 1.85 (*High*) shown in panel (c), respectively.

Figure 7 shows the predicted cumulative dust impacts produced by the dynamic impact flux model. Each orbit group and perihelion are color-coded, while the collected punctured events are marked on top of orbit numbers. Two curves represent the range of model predictions bounded by ISƟIS/EPI-Lo being impacted by 5 to 7 $\alpha$–meteoroids, scaling up by the corresponding factors of 1.42 and 1.85 that determined and shown in Fig. 6. Considering various spacecraft attitude (SPICE kernels) and orbital geometry (spacecraft attitude/pointing, Sun, and Earth), several orbits have higher possibilities to suffer dust hits before perihelion, for example, orbits 7 and 8.

We notice that the data-model comparison is quite consistent until the spacecraft entered the fourth orbit group (orbit 8 & 9 with perihelion at 16 Rs), where a total of four events punctured ISƟIS/EPI-Lo. For this small number of statistics encountered by ISƟIS, two assumptions have been considered: (a) ISƟIS impacts are representative of the average total impact rate PSP is encountering; thus, we normalize the model to detection results, or (b) ISƟIS may have encountered a less likely, but still within Poisson statistics, of more than average number of punctures given the impact rate estimated by PSP/FIELDS antenna instrument (uses entire spacecraft body as effective detection area). So far, the applied model has demonstrated its capability and robustness in predicting dust impact rates for scientific study and hazardous assessment purposes.

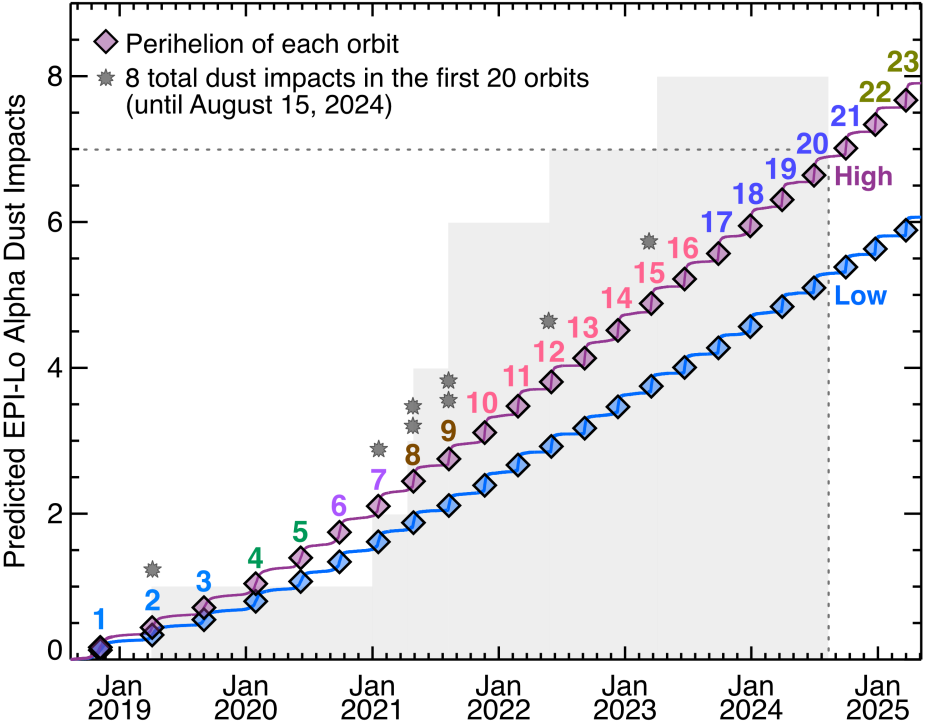

**Fig. 7:** Predicted $\alpha$–meteoroid dust hits curve from dynamic impact flux model and scale it up to match 5 or 7 $\alpha$–meteoroid impacts at the end of Orbit 20, respectively. Color indicates PSP's distinct orbital groups and shaded grey shows the cumulative eight total impacts.

# 5. Discussions and Conclusions

Throughout the first twenty orbits, ISʘIS has not only provided critical measurements of and insights into solar energetic particles but also registered direct dust detections within the intense dust environment of the inner heliosphere. Eight direct dust puncture events were recorded by penetrating telescope foils of EPI-Lo, thus enhancing the background contamination in eight of the eighty EPI-Lo apertures. Five of eight dust impacts occurred within 40 R$_S$ (0.19 au). While we are not able to precisely estimate the masses and speeds of these impacts, the known directionality allows us to make comparisons with a two-component dynamical model to assess the parent population for six impacts by comparing maps in Fig. 3 – 5.

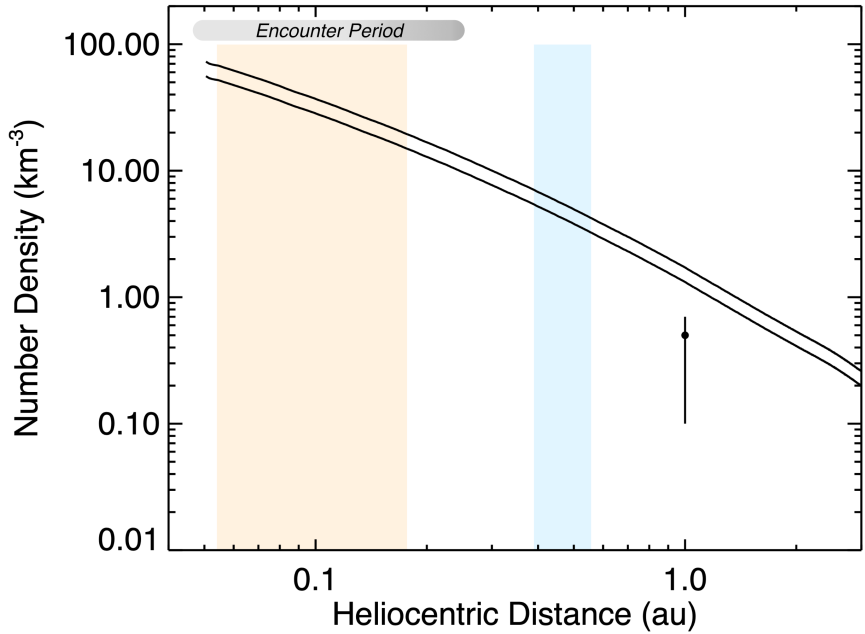

**Fig. 8:** Number density of $\alpha$–meteoroid in heliocentric distance derived from the dynamic impact flux model. Two bands indicate the range of puncture events ISʘIS encountered (five in light orange and three in light blue, cf. Fig.3), while two solid curves represent 0.64 – 0.89 μm.

Scaling the dynamical model from the analyses in this study, Fig. 8 shows the modeled number density of $\alpha$–meteoroids in the inner heliosphere. The single prior data point with errors is the 1 μm density at 1 au derived by *Szalay et al.* (2021), processed through PSP/FIELDS antenna dust measurements. It combines all raw detections with various heliocentric distances and is normalized back to 1 au. There is approximately a factor of 3 discrepancy between our derived number densities at 1 au and those derived from the PSP/FIELDS. However, the current model has a minimum particle size constraint of diameter D = 10 μm; anything below is an extrapolation that assumes everything D < 10 μm behaves the same dynamically. Recent analyses with PSP/FIELDS detections suggest the power-law size distribution for micron-sized grains near the Sun may be significantly steeper than expected such that there are many more grains with smaller radii than large grains (*Szalay, Pokorný, & Malaspina*, 2024). Hence, this discrepancy may very well be due to the difference between modeled and actual size distributions for micron-sized grains. Additionally, the single derived value shown here from *Szalay et al.* (2021) was extrapolated from measurements primarily inside 0.3 au and assumed the submicron grains also followed a density distribution proportional to $r^{1.3}$, which has not been directly verified for such small grains.

Of the eight direct impacts, we are able to identify the onset time and source of six impactors with five $\alpha$–meteoroids and one $\beta$–meteoroid via a data-model comparison of spacecraft location, impact flux, and impact speed. Of the five bound $\alpha$–meteoroid impactors, four were prograde, and one is consistent with a grain on a retrograde orbit with large impact speed (L28 event in orbit 9), hinting at the survival probability of long-period cometary material (including Oort Cloud Comets and Halley Type Comets) within 20 $R_S$.

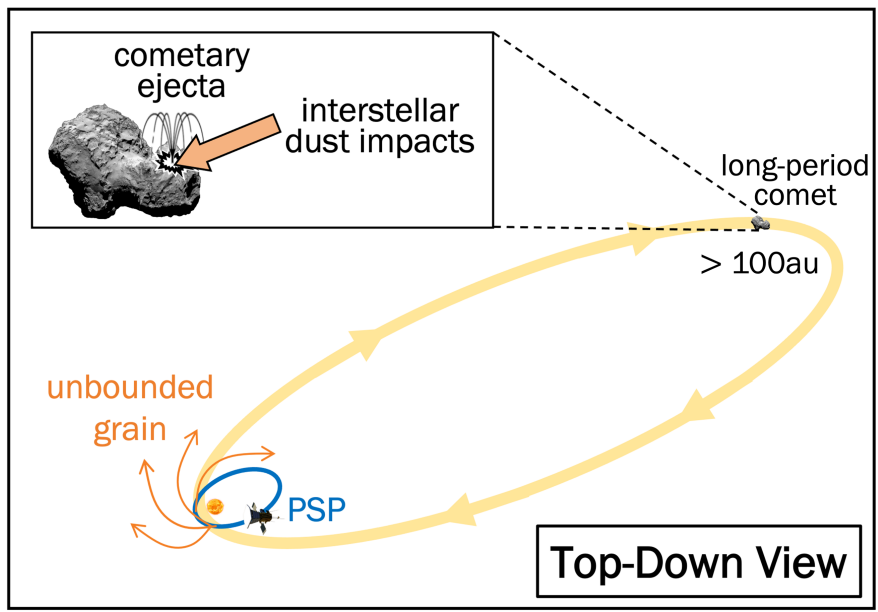

**Fig. 9:** Schematic diagram of retrograde grain shed from long-period cometary material that impacted ISʘIS/EPI-Lo instrument on Parker Solar Probe.

To achieve a high-speed retrograde impact of $\alpha$–meteoroids on ISƟIS/EPI-Lo, we suggest a scenario of a dust grain being expelled or shed from a "long-period comet," possibly due to an interstellar dust impact, as Fig. 9 illustrates. This would transform it into a retrograde $\alpha$–meteoroid, Following the parent comet's orbital motion. Cruising hundreds of years, it orbited toward the inner heliosphere, survived the dense collisional region near the Sun, and encountered the PSP/ISƟIS instrument. This grain cannot be too heavy/large (> 10s μm) to survive until impacting PSP, considering the collisional fragmentation process in the ambient interplanetary dust complex. Meanwhile, it cannot be too light/small (< 100s nm), or it would be expelled by radiation pressure before impact. This suggests long-period cometary materials may survive for significant amounts of time in the extreme near-sun environment.

Comets from the Oort Cloud originate at the extreme edge of the Solar System, representing remnants from the formation of the Solar System (*Oort*, 1950). Periodically, gravitational influences from passing stars and galactic tides can disturb an object from the Oort Cloud, causing its orbit to bring it closer to the Sun with a perihelion within the planetary region of the Solar System (< 30 au) (*Torres et al.,* 2019; *Vokrouhlický et al.*, 2019). Typically, the dust model predicts JFC comet grains dominate the interplanetary dust grain mass flux inside 10 au, while Oort-Cloud cometary grains and Edgeworth-Kuiper Belt grains govern the regions between 10 and 25 au and beyond 25 au, respectively (*Poppe*, 2016). Oort cloud cometary material is expected to exist, albeit in much lower abundance than from JFCs, in the inner heliosphere (*Campbell-Brown*, 2008; *Janches et al.*, 2014; *Nesvorný et al.,* 2010). OCCs that experience gravitational perturbation may alter their perihelion distances. Subsequently, a small subset of them may plunge close to the Sun and undergo intense cometary activity and/or destruction. Long-period comets often exhibit significant orbital inclinations, and a considerable number of them follow "retrograde" motion, many with orbital periods greater than 200 years (e.g., *Weissman*, 1997; *Fernández,* 2003; *Vokrouhlický et al.*, 2019). As previous literature stated, the physical disruption of OCCs is expected to result in significant releases of cometary material into the Solar System (*Nesvorný et al.,* 2010). Although a portion of this material may be expelled by solar radiation pressure, some could remain part of the interplanetary dust complex. Applying satellite observations of zodiacal cloud brightness, the contribution of OCCs is constrained to be ~10% of the interplanetary dust complex, which the uncertainty may vary but remains a nonnegligible population (*Nesvorný et al.,* 2010).

With the constraint of a "retrograde" grain impacting PSP, we consider OCCs the primary source of these grains due to their typical retrograde orbits. Additionally, the HTC dust population may have an imprint comparable to OCCs in the inner heliosphere bounded by observations. (*Pokorný et al.*, 2014; *Carrillo-Sanchez et al.*, 2016). However, the mixing ratios of HTCs and OCCs are not well-constrained (*Pokorný et al.*, 2019; *Carrillo-Sanchez et al.*, 2020).

In conclusion,

- PSP/ISⵙIS has registered eight direct dust puncture events for the first twenty orbits in the near Sun environment from August 2018 to August 2024, and several were the closest direct dust detections to the Sun.

- Investigating dust puncture events helps constrain the existing meteoroid model by providing location, flux, speed, and source category information.

- One of these impactors was likely from a retrograde dust grain, suggesting long-period cometary material may survive within 0.095 au (20.5 Rs).

- The data-model comparison with improved geometric factor and probability analysis demonstrates a well-constrained framework for future instrumental hazardous study.

# Appendix A. Example of coincidence measurements by PSP/FIELDS

Hypervelocity impacts ($\gtrsim$ 1km/s) upon the spacecraft body will produce an impact plasma cloud that generates transient voltage perturbations detected by antenna instruments (see *Shen et al.*, 2021b and references therein). On PSP, we leverage FIELDS electric field antennae (*Bale et al.*, 2016) to study interplanetary dust population and dynamics (e.g., *Szalay et al.*, 2020; 2021; *Page et al.*, 2020; *Malaspina et al.*, 2020; *Pusack et al.*, 2021). We deduce dust impacts on or near the ISⓄIS suite to produce measurable signals for PSP/FIELDS.

Searching for correlated impact signatures, we analyze bandpass filter (BPF) peak data from the FIELDS digital fields board (*Malaspina et al.*, 2016). The BPF peak data record the largest amplitude signal detected in each ~0.89s window on dipole channel V34, as V3 – V4. Fifteen bandpass frequency bins with center frequencies spanning ~0.4 Hz to 7 kHz (DC channel) and seven bandpass frequency bins with center frequencies spanning ~879 Hz to 50 kHz (AC channel) are examined for individual ISⓄIS/EPI-Lo impact events. Considering the time resolution of ISⓄIS/EPI-Lo impact determination is ~1 minute, the FIELDS/BPF data ±1 minute from each identified ISⓄIS impact is attached.

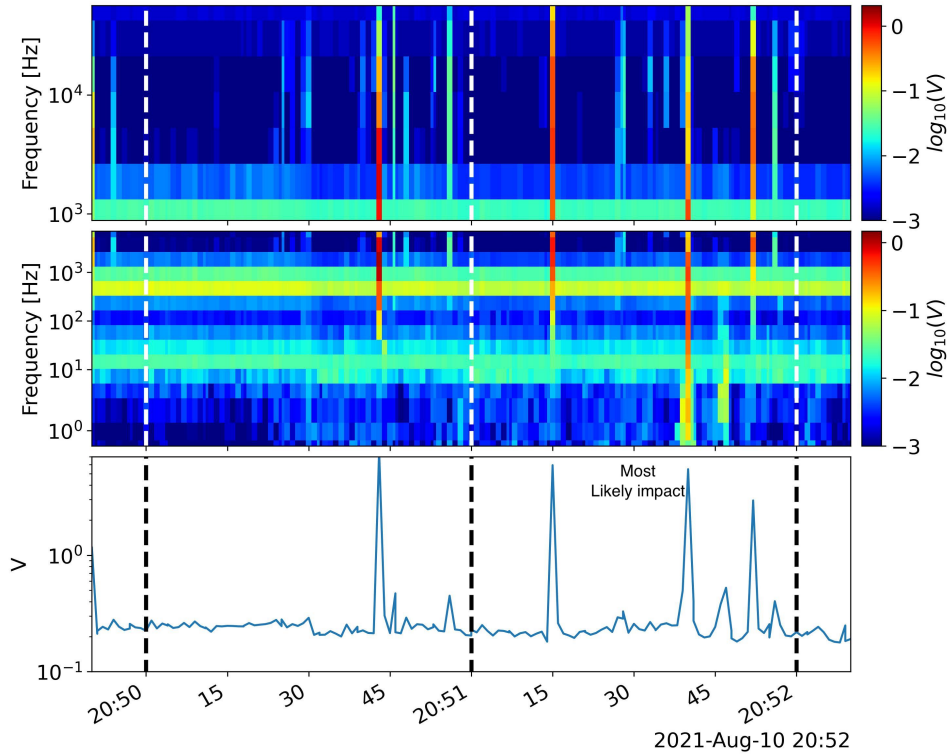

**Fig. A1:** Example PSP/FIELDS waveform of the most likely onset time of L28 dust puncture on orbit 9 based according to 2021–08–10 (222) 20:51 determined by PSP/ISⓄIS/EPI-Lo.

Figure A1 is a distinct example. The top and middle panels show the AC BPF peak and DC BPF peak data, respectively. The bottom panel shows the integral of BPF amplitudes at each time across both AC and DC frequency bins. Because dust impact voltage spikes are high amplitude transient impulses (< 1ms), they have broad frequency content, appearing in bandpass data as vertical lines, while in the integrated bandpass data as high amplitude spikes. The three vertical dashed lines indicate the -1 minute (left), +0 minutes (middle), and +1 minute from the ISƟIS identified impact time, 2021–08–10 (222) 20:51. During the two-minute interval, four prominent dust impacts are identified. We suggest the third one as the most likely impact to have punctured the ISƟIS start foil because the frequency content extends < 10 Hz. This indicates an impact waveform that persists longer than a typical dust impact voltage signal. Long-duration dust impact signatures are strongly correlated with the production of spacecraft debris clouds and suggest that the impacting dust grain was larger or faster than usual dust impacts (*Malaspina et al.*, 2022).

Of the eight ISƟIS foil impacts identified, the FIELDS instrument was powered on for five (L31, L21, L39, L28, and L56). In each of these five cases, dust impacts were observed during the ±1 min period about each ISƟIS foil impact. For L31, a single dust impact is observed during the ±1 min window. It extends to low (< 10 Hz) frequencies, so we identify this as the most likely impact event. All other intervals show multiple impact signatures, which is not surprising given the high flux of dust impacts experienced by PSP (*Malaspina et al.*, 2023). Dust impacts that reach < 10 Hz are also observed during the L39 and L28 ±1 min windows. The L21 and L56 events show > 6 impact signatures in their ±1 min windows. Any one of these could be a signature of an ISƟIS foil impact, but none extend < 10 Hz. We generally observe correlations between dust puncture on PSP/ISƟIS and transient voltage spikes measured by PSP/FIELDS. However, the exact correlation is restricted by different instrument cadences.

# Acknowledgment

We gratefully thank everyone who helped make the ISΘIS instrument suite and PSP mission possible. The ISΘIS data and visualization tools are available to the community at https://spacephysics.princeton.edu/missions-instruments/PSP; data are also available via the NASA Space Physics Data Facility (https://spdf.gsfc.nasa.gov/) and ISΘIS science operation center (SOC, https://spp-isois.sr.unh.edu/home.html). Parker Solar Probe was designed, built, and is operated by the Johns Hopkins Applied Physics Laboratory (APL) as part of NASA's Living with a Star (LWS) program (contract NNN06AA01C). Support from the LWS management and technical team has played a critical role in the success of the Parker Solar Probe mission. The modeling material is based upon work supported by NASA under awards number 80GSFC24M0006, 80NSSC21K0153, and ISFM work packages EIMM and Planetary Geodesy at NASA Goddard Space Flight Center. We sincerely appreciate Carol Weaver's ISΘIS SOC role in providing first-glance dust impact alerts and attitude kernels for simulations in coordination with the PSP mission operation center at APL.

# Data Availability Statement

The data (Shen et al., 2024, Supporting Information data set) are publicly available in the Zenodo repository (https://doi.org/10.5281/zenodo.14548706).

# Author contributions (if required)

Conceptualization – J. R. Szalay, M. M. Shen

Formal analysis – M. M. Shen, J. R. Szalay, P. Pokorný, D. Malaspina

Funding acquisition - J. R. Szalay, P. Pokorný, D. Malaspina

Investigation –  M. M. Shen, J. R. Szalay

Methodology –J. R. Szalay, M. M. Shen

Validation – M. M. Shen, J. R. Szalay, P. Pokorný, M. E. Hill, J. G. Mitchell, D. Malaspina

Writing – original draft –  M. M. Shen, J. G. Mitchell, J. R. Szalay

Writing – review & editing – M. M. Shen, J. R. Szalay, J. G. Mitchell, P. Pokorný, D. J. McComas, D. Malaspina, C. M. S. Cohen

# References

Agostinelli, S., Allison, J., Amako, K. A., Apostolakis, J., Araujo, H., Arce, P., ... & Geant4 Collaboration. (2003). GEANT4—a simulation toolkit. Nuclear instruments and methods in physics research section A: Accelerators, Spectrometers, Detectors and Associated Equipment, 506(3), 250-303.

Allen, R. C., Lario, D., Odstrcil, D., Ho, G. C., Jian, L. K., Cohen, C. M. S., ... & Wiedenbeck, M. (2020). Solar wind streams and stream interaction regions observed by the parker solar probe with corresponding observations at 1 AU. The Astrophysical Journal Supplement Series, 246(2), 36.

Auer, S., (2001). Instrumentation, in Interplanetary Dust, edited by E. Grün, pp. 385–444, Springer, New York.

Bale, S.D., Goetz, K., Harvey, P.R. et al. The FIELDS Instrument Suite for Solar Probe Plus. Space Sci Rev 204, 49–82 (2016). https://doi.org/10.1007/s11214-016-0244-5

Berg, O., and E. Grün (1973), Evidence of hyperbolic cosmic dust particles., in Plenary Meeting on Space research XIII.

Burns, J. A., Lamy, P. L., & Soter, S. (1979). Radiation forces on small particles in the solar system. Icarus, 40(1), 1-48.

Campbell-Brown, M. D. (2008). High resolution radiant distribution and orbits of sporadic radar meteoroids. Icarus, 196(1), 144-163.

Carrillo-Sánchez, J. D., D. Nesvorný, P. Pokorný, D. Janches, and J. M. C. Plane (2016), Sources of cosmic dust in the Earth's atmosphere, Geophys. Res. Lett., 43, 11,979–11,986, doi:10.1002/2016GL071697.

Carrillo-Sánchez, J. D., Gómez-Martín, J. C., Bones, D. L., Nesvorný, D., Pokorný, P., Benna, M., ... & Plane, J. M. (2020). Cosmic dust fluxes in the atmospheres of Earth, Mars, and Venus. Icarus, 335, 113395.

Cohen, C. M. S., Christian, E. R., Cummings, A. C., Davis, A. J., Desai, M. I., Giacalone, J., ... & Whittlesey, P. (2020). Energetic particle increases associated with stream interaction regions. The Astrophysical Journal Supplement Series, 246(2), 20.

Cukier, W. Z., & Szalay, J. R. (2023). Formation, Structure, and Detectability of the Geminids Meteoroid Stream. The Planetary Science Journal, 4(6), 109.

Desai, M. I., Mitchell, D. G., Szalay, J. R., Roelof, E. C., Giacalone, J., Hill, M. E., ... & Kasper, J. C. (2020). Properties of suprathermal-through-energetic He ions associated with stream interaction regions observed over the Parker Solar Probe's first two orbits. The Astrophysical Journal Supplement Series, 246(2), 56.

Fernández, J.A. (2002). Long-Period Comets and the Oort Cloud. In: Boehnhardt, H., Combi, M., Kidger, M.R., Schulz, R. (eds) Cometary Science after Hale-Bopp. Springer, Dordrecht. https://doi.org/10.1007/978-94-017-1086-2_15

Fox, N.J., Velli, M.C., Bale, S.D. et al. The Solar Probe Plus Mission: Humanity's First Visit to Our Star. Space Sci Rev 204, 7–48 (2016). https://doi.org/10.1007/s11214-015-0211-6

Grün, E., N. Pailer, H. Fechtig, and J. Kissel (1980), Orbital and physical characteristics of micrometeoroids in the inner solar system as observed by helios 1, Planetary and Space Science, 28 (3), 333 – 349, https://doi.org/10.1016/0032-0633(80)90022-7.

Grün, E., H. Zook, H. Fechtig, and R. Giese (1985), Collisional balance of the meteoritic complex, Icarus, 62 (2), 244 – 272, https://doi.org/10.1016/0019-1035(85)90121-6.

Hill, M. E., Mitchell, D. G., Andrews, G. B., Cooper, S. A., Gurnee, R. S., Hayes, J. R., ... & Westlake, J. H. (2017). The Mushroom: A half-sky energetic ion and electron detector. Journal of Geophysical Research: Space Physics, 122(2), 1513-1530.

Ishimoto, H., and I. Mann (1998), Modeling the particle mass distribution within 1 AU of the Sun, Planetary and Space Science, 47 (1), 225 – 232.

Janches, D., Plane, J. M. C., Nesvorný, D., Feng, W., Vokrouhlický, D., & Nicolls, M. J. (2014). Radar detectability studies of slow and small zodiacal dust cloud particles. I. The case of Arecibo 430 MHz meteor head echo observations. The Astrophysical Journal, 796(1), 41.

Jenniskens, P. (1994), Meteor stream activity I. The annual streams, Astronomy and Astrophysics 287, 287, 990 – 1013.

Jewitt, D., Li, J., & Agarwal, J. (2013). The dust tail of asteroid (3200) Phaethon. The Astrophysical Journal Letters, 771(2), L36.

Kasper, J.C., Abiad, R., Austin, G. et al. Solar Wind Electrons Alphas and Protons (SWEAP) Investigation: Design of the Solar Wind and Coronal Plasma Instrument Suite for Solar Probe Plus. Space Sci Rev 204, 131–186 (2016). https://doi.org/10.1007/s11214-015-0206-3

Malaspina, D. M., Ergun, R. E., Bolton, M., Kien, M., Summers, D., Stevens, K., ... & Goetz, K. (2016). The Digital Fields Board for the FIELDS instrument suite on the Solar Probe Plus mission: Analog and digital signal processing. Journal of Geophysical Research: Space Physics, 121(6), 5088-5096.

Malaspina, D. M., Szalay, J. R., Pokorný, P., Page, B., Bale, S. D., Bonnell, J. W., ... & Pulupa, M. (2020). In situ observations of interplanetary dust variability in the inner heliosphere. The Astrophysical Journal, 892(2), 115.

Malaspina, D. M., Stenborg, G., Mehoke, D., Al-Ghazwi, A., Shen, M. M., Hsu, H. W., ... & de Wit, T. D. (2022). Clouds of spacecraft debris liberated by hypervelocity dust impacts on parker solar probe. The Astrophysical Journal, 925(1), 27.

Malaspina, D. M., A. Toma, J. R. Szalay, M. Pulupa, P. Pokorn´y, S. D. Bale, and K. Goetz (2023), A dust detection database for the inner heliosphere using the parker solar probe spacecraft, The Astrophysical Journal Supplement Series, 266 (2), 21, doi:10.3847/1538-4365/acca75.

Mann, I., and H. Kimura (2000), Interstellar dust properties derived from mass density, mass distribution, and flux rates in the heliosphere, Journal of Geophysical Research:Space Physics, 105 (A5), 10,317 – 10,328, https://doi.org/10.1029/1999JA900404.

Mann, I., Kimura, H., Biesecker, D.A. et al. Dust Near The Sun. Space Science Reviews 110, 269–305 (2004). https://doi.org/10.1023/B:SPAC.0000023440.82735.ba

Meyer-Vernet, N., M. Moncuquet, K. Issautier, and A. Lecacheux (2014), The importance of monopole antennas for dust observations: Why wind/waves does not detect nanodust, Geophysical Research Letters, 41 (8), 2716 – 2720, https://doi.org/10.1002/2014GL059988.

McComas, D.J., Alexander, N., Angold, N. et al. Integrated Science Investigation of the Sun (ISIS): Design of the Energetic Particle Investigation. Space Sci Rev 204, 187–256 (2016). https://doi.org/10.1007/s11214-014-0059-1

McComas, D.J., Christian, E.R., Cohen, C.M.S. et al. Probing the energetic particle environment near the Sun. Nature 576, 223–227 (2019). https://doi.org/10.1038/s41586-019-1811-1

Mitchell, J. G., Leske, R. A., Nolfo, G. D., Christian, E. R., Wiedenbeck, M. E., McComas, D. J., ... & Szalay, J. R. (2022). First Measurements of Jovian Electrons by Parker Solar Probe/IS⊙IS within 0.5 au of the Sun. The Astrophysical Journal, 933(2), 171.

Mitchell, J. G., de Nolfo, G. A., Christian, E. R., Leske, R. A., Ryan, J. M., Vievering, J. T., ... & Schwadron, N. A. (2024). IS⊙IS Solar γ-Ray Measurements: Initial Observations and Calibrations. The Astrophysical Journal, 968(1), 33.

Morfill, G. E., E. Gr¨un, and C. Leinert (1986), The interaction of solid particles with the interplanetary medium, in The Sun and the Heliosphere in Three Dimensions, edited by R. G. Marsden, pp. 455 – 474, Springer Netherlands, Dordrecht.

Mozer, F. S., Agapitov, O. V., Bale, S. D., Bonnell, J. W., Goetz, K., Goodrich, K. A., ... & Schumm, G. (2020). Time domain structures and dust in the solar vicinity: Parker solar probe observations. The Astrophysical Journal Supplement Series, 246(2), 50.

Nesvorný, D., Jenniskens, P., Levison, H. F., Bottke, W. F., Vokrouhlický, D., & Gounelle, M. (2010). Cometary origin of the zodiacal cloud and carbonaceous micrometeorites. Implications for hot debris disks. The Astrophysical Journal, 713(2), 816.

Nesvorný, D., Vokrouhlický, D., Dones, L., Levison, H. F., Kaib, N., & Morbidelli, A. (2017). Origin and evolution of short-period comets. The Astrophysical Journal, 845(1), 27.

Oort, J. H. 1950, Bull. Astron. Inst. Netherlands, 11, 91

Page, B., Bale, S. D., Bonnell, J. W., Goetz, K., Goodrich, K., Harvey, P. R., ... & Szalay, J. R. (2020). Examining dust directionality with the Parker solar probe FIELDS instrument. The Astrophysical Journal Supplement Series, 246(2), 51.

Pokorný, P., Vokrouhlický, D., Nesvorný, D., Campbell-Brown, M., & Brown, P. (2014). Dynamical model for the toroidal sporadic meteors. The Astrophysical Journal, 789(1), 25.

Pokorný, P., Sarantos, M., & Janches, D. (2018). A comprehensive model of the meteoroid environment around Mercury. The Astrophysical Journal, 863(1), 31.

Pokorný, P., & Kuchner, M. (2019). Co-orbital Asteroids as the Source of Venus's Zodiacal Dust Ring. The Astrophysical Journal Letters, 873(2), L16.

Pokorný, P., Mazarico, E., & Schorghofer, N. (2021). Erosion of Volatiles by Micrometeoroid Bombardment on Ceres and Comparison to the Moon and Mercury. The Planetary Science Journal, 2(3), 85.

Pokorný, P., Moorhead, A. V., Kuchner, M. J., Szalay, J. R., & Malaspina, D. M. (2024). How long-lived grains dominate the shape of the Zodiacal Cloud. The Planetary Science Journal, 5(3), 82.

Poppe, A. R. (2016). An improved model for interplanetary dust fluxes in the outer Solar System. Icarus, 264, 369-386.

Pusack, A., Malaspina, D. M., Szalay, J. R., Bale, S. D., Goetz, K., MacDowall, R. J., & Pulupa, M. (2021). Dust directionality and an anomalous interplanetary dust population detected by the Parker solar probe. The Planetary Science Journal, 2(5), 186.

Rankin, J. S., McComas, D. J., Leske, R. A., Christian, E. R., Cohen, C. M. S., Cummings, A. C., ... & Wiedenbeck, M. E. (2021). First observations of anomalous cosmic rays in to 36 solar radii. The Astrophysical Journal, 912(2), 139.

Rankin, J. S., McComas, D. J., Leske, R. A., Christian, E. R., Cohen, C. M. S., Cummings, A. C., ... & Wiedenbeck, M. E. (2022). Anomalous cosmic-ray oxygen observations into 0.1 au. The Astrophysical Journal, 925(1), 9.

Schwadron, N.A., Gloeckler, G. Pickup Ions and Cosmic Rays from Dust in the Heliosphere. Space Sci Rev 130, 283–291 (2007). https://doi.org/10.1007/s11214-007-9166-6

Shen, M. M., Sternovsky, Z., Horányi, M., Hsu, H. W., & Malaspina, D. M. (2021a). Laboratory study of antenna signals generated by dust impacts on spacecraft. Journal of Geophysical Research: Space Physics, 126(4), e2020JA028965.

Shen, Mitchell M., Sternovsky, Zoltan, Garzelli, Alessandro, & Malaspina, David M. (2021b). Supplementary data to: "Electrostatic model for antenna signal generation from dust impacts" [Data set]. Zenodo. http://doi.org/10.5281/zenodo.4888925

Shen, M. M., Sternovsky, Z., & Malaspina, D. M. (2023). Variability of antenna signals from dust impacts. Journal of Geophysical Research: Space Physics, 128, e2022JA030981. https://doi.org/10.1029/2022JA030981

Shen, M., Szalay, J., Pokorny, P., Malaspina, D., McComas, D., & Bale, S. (2024). Supplementary data to "Diverse dust populations in the near-Sun environment characterized by PSP/ISOIS" [Data set]. Zenodo. https://doi.org/10.5281/zenodo.14548706

Soja, R. H., Grün, E., Strub, P., Sommer, M., Millinger, M., Vaubaillon, J., ... & Srama, R. (2019). IMEM2: a meteoroid environment model for the inner solar system. Astronomy & Astrophysics, 628, A109.

Sommer, M. (2023). Alpha-Meteoroids then and now: Unearthing an overlooked micrometeoroid population. Planetary and Space Science, 236, 105751.

Szalay, J. R., Pokorný, P., Bale, S. D., Christian, E. R., Goetz, K., Goodrich, K., ... & McComas, D. J. (2020). The Near-Sun Dust Environment: Initial Observations from Parker Solar Probe. The Astrophysical Journal Supplement Series, 246(2), 27.

Szalay, J. R., Pokorný, P., Malaspina, D. M., Pusack, A., Bale, S. D., Battams, K., ... & Strub, P. (2021). Collisional evolution of the inner zodiacal cloud. The Planetary Science Journal, 2(5), 185.

Szalay, J. R., Pokorný, P., & Malaspina, D. M. (2024). Size distribution of small grains in the inner zodiacal cloud. The Planetary Science Journal, 5(12), 266.

Torres, S., Cai, M. X., Brown, A. G. A., & Zwart, S. P. (2019). Galactic tide and local stellar perturbations on the Oort cloud: creation of interstellar comets. Astronomy & Astrophysics, 629, A139.

Vokrouhlický, D., Nesvorný, D., & Dones, L. (2019). Origin and evolution of long-period comets. The Astronomical Journal, 157(5), 181.

Vourlidas, A., Howard, R.A., Plunkett, S.P. et al. The Wide-Field Imager for Solar Probe Plus (WISPR). Space Sci Rev 204, 83–130 (2016). https://doi.org/10.1007/s11214-014-0114-y

Wehry, A., and I. Mann (1999), Identification of $\beta$-meteoroids from measurements of the dust detector onboard the ULYSSES spacecraft, A&A, 341, 296–303.

WEISSMAN, P.R. (1997), Long-period Comets and the Oort Cloud. Annals of the New York Academy of Sciences, 822: 67-95. https://doi.org/10.1111/j.1749-6632.1997.tb48335.x

Wimmer-Schweingruber, R. F., and P. Bochsler (2003), On the origin of inner-source pickup ions, Geophys. Res. Lett., 30, 1077, doi:10.1029/2002GL015218, 2.

Zaslavsky, A., Meyer-Vernet, N., Mann, I., Czechowski, A., Issautier, K., Le Chat, G., ... & Kasper, J. C. (2012). Interplanetary dust detection by radio antennas: Mass calibration and fluxes measured by STEREO/WAVES. Journal of Geophysical Research: Space Physics, 117(A5).

Zaslavsky, A., Mann, I., Soucek, J., Czechowski, A., Píša, D., Vaverka, J., ... & Vaivads, A. (2021). First dust measurements with the Solar Orbiter Radio and Plasma Wave instrument. Astronomy & Astrophysics, 656, A30.

Zook, H. A., and O. E. Berg (1975), A source for hyperbolic cosmic dust particles, Planetary and Space Science, 23 (1), 183–203, https://doi.org/10.1016/0032-0633(75)90078-1.